\documentclass[12pt]{article}
\usepackage{graphicx}
\usepackage[bookmarksnumbered,colorlinks,plainpages]{hyperref}

\begin{document}

\title{Maximum path information and Fokker-Planck Equation}

\author{W. Li$^{\dag\ddag}$\thanks{Corresponding author. Electronic address: liw@phy.ccnu.edu.cn. Tel: 86-27-6786-7046.},
Q.A. Wang$^{\ddag}$, A. Le M\'ehaut\'e$^{\ddag}$\\
{\small $^{\dag}$Complexity Science Center and Institute of Particle Physics}, \\
{\small Hua-Zhong Normal University, Wuhan 430079, P.R. China}\\
{\small $^{\ddag}$Institut Sup\'erieur des Mat\'eriaux du Mans}, \\
{\small 44, Avenue F.A. Bartholdi, 72000 Le Mans, France}}

\date{}

\maketitle

\begin{abstract}

We present in this paper a rigorous method to derive the nonlinear
Fokker-Planck (FP) equation of anomalous diffusion directly from a
generalization of the principle of least action of Maupertuis
proposed by Wang \cite{Wang1} for smooth or quasi-smooth irregular
dynamics evolving in Markovian process. The FP equation obtained may
take two different but equivalent forms. It was also found that the
diffusion constant may depend on both q (the index of Tsallis
entropy \cite{Tsallis1}) and the time t.

\end{abstract}

The Fokker-Planck equation is a differential equation describing
the time evolution of probability distribution of state during
stochastic processes. The FP equation and its generalizations play
very crucial roles in statistical physics. The FP equation is not
only applicable to the systems near the thermal equilibrium, but
to the systems far from the thermal equilibrium as well. This
latter application has special meaning in dealing with a large
class of self-organized, complex dynamical systems. In this sense,
the FP equation not only describes stationary properties but also
the dynamics of evolving systems.

The FP equation was first derived by Fokker \cite{Fokker} and
\cite{Planck} as one to describe Brownian motion. Later on, many
books and review articles were published \cite{Uhlenbeck,
Chandrasekhar, Wang&Uhlenbeck, Haken, Schuss}. The usual way of
deriving the FP equation starts from calculating the transition
probability $P(x,t+\tau|x^{\prime},t)$ for small $\tau$, where the
particle travels from $x$ to $x^{\prime}$. There are various ways
to derive the expression of the transition probability
\cite{Risken}. In this paper, we employ the concept of maximum
path information, related to non-extensive Tsallis entropy, to
derive the expression of the transition probability for the motion
of particle under the influence of external forces. On the basis
of the transition probability, an nonlinear FP equation can be
obtained.

A path information based on Shannon entropy \cite {Shannon1} has
been defined as \cite{Wang1}
\begin{equation}
\label{shannon1}
 H_{s}(a,b)=-\sum_{i=1}^{w} p_{ab}(i) \ln
p_{ab}(i),
\end{equation}
\noindent where $p_{ab}(i)$ is called the transition probability
that a system moving from point $a$ to point $b$ will choose path
$i$ among all possibly existed $w$ paths. In the case of
non-extensive statistics, the corresponding path information
naturally adopts the formula of Tsallis entropy \cite{Tsallis1}
\begin{equation}
\label{tsallis1}
 H_{t} (a,b)=-k \sum_{i=1}^{w} \frac
{p_{ab}(i)-p^q_{ab}(i)}{1-q},
\end{equation}
\noindent where $q$, the entropy index, specifies a particular
statistics. In general, the larger the path information, the less
we know about paths states of the system.

Of course, the transition probability $p_{ab} (i)$ satisfies the
following normalization condition
\begin{equation}
\label{normalization} \sum_{i=1}^{w} p_{ab}(i)=1.
\end{equation}
For classical dynamical systems we also suppose each possible path
is characterized by its action $A_{ab}(i)$
\begin{equation}\label{action}
A_{ab}(i)=\int_{t_{ab}} L_{i} (t) dt,
\end{equation}
\noindent where $L_{i} (t)$ is the Lagrangian of the system at
time $t$ via the path $i$. The average action is represented by
\begin{equation}
\label{averageaction} \langle A_{ab} \rangle = \sum_{i=1}^{w}
A_{ab} (i) p_{ab} (i).
\end{equation}

In order to obtain the form of path probability, we seek to
optimize the path information $H_{t} (a,b)$ under the constraints
of Eqs. (\ref{normalization}) and (\ref{averageaction}). That is,
\begin{equation}
\label{maximum} \delta (H_t(a,b)+\alpha \sum_{i=1}^{w}
p_{ab}(i)+\eta \sum_{i=1}^{w} A_{ab} (i) p_{ab} (i))=0
\end{equation}
Through a simple algebra, the optimization yields the following
expression of path probability
\begin{equation}\label{path-probability}
p_{ab} (i)=\frac {1}{Z_q} [1-(1-q)\eta A_{ab}(i)]^{\frac {1}
{1-q}},
\end{equation}
\noindent where $Z_q=\sum_{i=1}^{w} [1-(1-q)\eta A_{ab}(i)]^{\frac
{1} {1-q}}$.

In order to obtain a general derivation of FP equation at the
existence of any form of external forces (drifts), we adopt here the
Euler's method to calculate the action. The detailed method is as
follows. The path through which the particle travels from point $a$
to point $b$ is cut into $N$ segments with each having a spatial
length $\Delta x_k=(x_k-x_{k-1})$ ($k=1,2,...,N$). $t=t_k-t_{k-1}$
is the time interval spent by the system on every segment. According
to the theorem of large numbers, the fluctuation of calculation will
go to 0 as $N$ approaches infinity. The action $A_k$ on the segment
$k$ is simply
\begin{equation}\label{actionsegment}
A_k=\frac {m(\Delta x_k)^2} {2t}+\frac{\Delta
x_k}{2}F_k-U(x_{k-1})t,
\end{equation}
\noindent where $F_k=-(\frac {\partial U}{\partial x})_k$ and
$U(x_{k-1})$ is the potential energy at the point $x_{k-1}$. Here
in this paper $F_k$ and $U(x_{k-1})$ will be considered as
constant. From now on, we will write $U(x_{k-1})$ as $U$ for
simplicity.

By using Eq. (\ref{path-probability}) the transition probability
$p_{k/k-1}$ from point $k-1$ to point $k$ via the path $i$ can be
written as
\begin{eqnarray}\label{transition1}\nonumber
p_{k/k-1}(i) &=&
\frac {1} {Z_q (k,k-1)} \{1-(1-q)\eta[\frac{m(\Delta x_k)^2}{2t}\\
&&+\frac {F_k t} {2} \Delta x_k-Ut]\}^{\frac {1}{1-q}},
\end{eqnarray}
\noindent where $Z_q(k,k-1)$ can be calculated from the
normalization condition
$\int_{-\infty}^{+\infty}p_{k/k-1}(i)dx_k=1$
\begin{eqnarray}\label{Zq1}\nonumber
Z_q(k,k-1)&=&\int_{-\infty}^{+\infty}dx_k\{1-(1-q)\eta[\frac{m(\Delta
x_k)^2}{2t}\\&&+\frac {F_k t} {2} \Delta x_k-Ut]\}^{\frac
{1}{1-q}}.
\end{eqnarray}

Introducing the methods by Tsallis and Prato \cite{Tsallis2,
Prato}, after a tedious calculation, we obtain the exact form of
$Z_q(k,k-1)$
\begin{equation}\label{Zq2}
Z_q(k,k-1)=A(q) \sqrt{\frac{2 \pi t} {\eta m}}
[1-(q-1)\eta(\frac{F_k^2t^3}{8m}+Ut)]^{\frac{q-3}{2q-2}},
\end{equation}
\noindent where $A(q)$ can be written as
\[A(q)=\left\{ \begin{array}{r@{\quad,\quad}l} \Gamma^{-1}(\frac{1}{q-1})
\Gamma(\frac{q-3}{2q-2})\sqrt{\frac{1}{q-1}}& q
>1 ;\\ \Gamma (\frac{2-q}{1-q})
\Gamma^{-1}(\frac{3-q}{2-2q})\sqrt{\frac{1}{1-q}}& 0<q<1.
\end{array}
\right. \] \noindent It is not difficult to prove that
$Z_q(k,k-1)$ restores to $Z_1(k,k-1)$ at the $q \rightarrow 1$
limit, which is
\begin{equation}
\label{Z_1}
Z_1(k,k-1)=exp[\eta(\frac{F_k^2t^3}{8m}+Ut)]\sqrt{\frac{2\pi t}
{\eta m}}.
\end{equation}

\noindent Hence the transition probability $p_{k/k-1}(i)$ has the
form
\begin{eqnarray}\label{transition2}\nonumber
p_{k/k-1}(i) &=&B(q)t^{-1/2}[1-(q-1)\eta
(\frac{F^2_kt^3}{8m}+Ut)]^{\frac{3-q}{2q-2}}\{1-\\&&(1-q)\eta
 [\frac{m(\Delta
x_k)^2}{2t}+\frac {F_k t} {2} \Delta x_k-Ut]\}^{\frac {1}{1-q}},
\end{eqnarray}
\noindent where $B(q)=A^{-1}(q)\sqrt{\frac{m\eta}{2\pi}}$.

Now we are ready to derive the FP equation for the system
travelling through the $k$-th segment of path $i$ connecting
points $a$ and $b$. It is readily that
\begin{eqnarray}\label{partialT}\nonumber
 \frac {\partial p_{k/k-1}(i)} {\partial t}&=& p_{k/k-1}(i)
  \{-\frac{1}{2}t^{-1}+A^{-1}_1 \frac{(q-3)\eta}{2}(\frac{3F^2_kt^2}{8m}+\\ && U)
 +A^{-1}_2 \eta [\frac {m}{2t^2}(\Delta x_k)^2-\frac{F_k \Delta x_k}{2}+U]\},
\end{eqnarray}
\noindent where $A_1=1-(q-1)\eta (\frac{F^2_kt^3}{8m}+Ut)$, and
$A_2=1-(1-q)\eta[\frac{m}{2t}(\Delta x_k)^2+\frac{F_k \Delta
x_k}{2}t-Ut]$. \noindent We also have
\begin{equation}\label{partialX1}
\frac {\partial p_{k/k-1}(i)} {\partial x_k}=
-p_{k/k-1}(i)A^{-1}_2\eta(mt^{-1}\Delta x_k+\frac{F_kt}{2}),
\end{equation}
\noindent and
\begin{eqnarray}\label{partialX2}\nonumber
\frac {\partial^2 [p_{k/k-1}(i)]^{\gamma}}{\partial
x_k^2}&=&-\gamma \eta A^{-1}_2
[p_{k/k-1}(i)]^{\gamma}[mt^{-1}-(\gamma-1+q)\\&& \times \eta
A^{-1}_2(mt^{-1}\Delta x_k+\frac{F_kt}{2})^2],
\end{eqnarray}
\noindent where $\gamma$ is a constant that might depend on $q$.

Combining the equations (\ref{partialT}) and (\ref{partialX1}),
one obtains the following expression
\begin{equation}\label{FPleft}
(\frac {\partial}{\partial t}+F_k \frac {\partial} {\partial x_k})
p_{k/k-1}(i)=-\frac {p_{k/k-1}(i)}{2m}[u_1(x_k,t,q)+v(x_k,t,q)],
\end{equation}
\noindent where
\begin{equation} u_1(x_k,t,q)=mt^{-1}-\eta
A^{-1}_2(mt^{-1}\Delta x_k+\frac {F_k t}{2})^2
\end{equation} and
\begin{eqnarray}\nonumber
v(x_k,t,q)&=&(\frac {3F^2_k t^2}{8m}+U)(3-q)\eta m A^{-1}_1 +2\eta
mA^{-1}_2(\frac{F^2_kt}{2}\\&& +\frac{F^2_kt^2}{8m}+\frac{mF_k
\Delta x_k}{t}+F_k\Delta x_k-U).
\end{eqnarray}

Writing (\ref{partialX2}) in another form one gets
\begin{equation}\label{FPright}
\frac {\partial^2 [p_{k/k-1}(i)]^{\gamma}}{\partial x_k^2}=-\gamma
\eta A^{-1}_2 [p_{k/k-1}(i)]^{\gamma}u_2(x,t,q,\gamma),
\end{equation}
\noindent where
\begin{equation}
u_2(x_k,t,q,\gamma)=mt^{-1}-(\gamma-1+q)\eta
A^{-1}_2(mt^{-1}\Delta x_k+\frac{F_kt}{2})^2.
\end{equation}
\noindent It is obvious that $u_2(x_k,t,q,2-q)=u_1(x_k,t,q)$.

Relating Eqs.(\ref{FPleft}) and (\ref{FPright}), together with
$u_2(x_k,t,q,2-q)=u_1(x_k,t,q)$, one obtains the following
equation
\begin{equation}\label{FP}
(\frac {\partial}{\partial t}+F_k \frac {\partial} {\partial x_k})
p_{k/k-1}(i)=D(q,t)\frac {\partial^2
[p_{k/k-1}(i)]^{2-q}}{\partial x_k^2},
\end{equation}
where
\begin{equation}\label{D(q,t)}
D(q,t)=\frac{1}{2\eta
m}B^{q-1}A^{(3-q)/2}_1t^{(1-q)/2}[1-\frac{v(x_k,t,q)}{u_1(x_k,t,q)(2-q)}].
\end{equation}
\noindent One can check that $D(1,t)=1/2\eta m$, which is
consistent with the results in \cite{Wang1}. Apparently, Eq.
(\ref{FP}) is the exact FP equation for the system in an
infinitesimally interval in the existence of external forces.

Besides Eq. (\ref{FP}), the FP equation can also take another
form,
\begin{eqnarray}\label{anotherFP}\nonumber
\frac{\partial}{\partial t}
[p_{k/k-1}(i)]^{2-q}&=&-\frac{\partial}{\partial
x_k}\{F_k[p_{k/k-1}(i)]^{2-q})\}+D^{\prime}(q,t)\\
&& \times \frac{\partial^2}{\partial x_k^2}[p_{k/k-1}(i)]^{2-q},
\end{eqnarray}
\noindent where
\begin{equation}
D^{\prime}(q,t)=\frac{A_2}{2\eta m}
[1-\frac{v(x_k,t,q)}{u_1(x_k,t,q)(2-q)}].
\end{equation}

We note that $D(q,t)$ in Eq. (\ref{FP}) and $D^{\prime}(q,t)$ in Eq.
(\ref{anotherFP}) are both $q$ and $t$ dependent. The dependence on
$q$ is a direct consequence of the non-extensive statistics where
$q$ is the identity of the system described. It has been shown above
that when $q=1$, the normal diffusion constant can be restored. The
dependence on $t$ is also quite natural because we are now dealing
with evolutionary processes where the phase space through which the
diffusion occurs is changing with time. As $t \rightarrow \infty$,
one readily obtains the diffusion constant for the stationary state.

The nonlinear FP equation derived above, Eq. (\ref{FP}) and Eq.
(\ref{anotherFP}), is well applied to describing the evolutionary
processes and stochastic processes of a large class of
self-organized systems that are far from thermal equilibrium, as
well as chemical equilibrium, such as transportation and diffusion
occurred in fractal or curved space. For example, it can be employed
to describe the broad range of markets and exchanges characterized
by the anomalous (super) diffusion and power-law distributions
\cite{Michael}. Another hope is that this equation can be applied to
the complex biological systems where evolution and anomalous
diffusion are taking place from time to time. Compared to the normal
FP equation and some of its other nonlinear forms \cite{Tsallis3},
our FP equation is more general because it can describe both regular
dynamics and irregular dynamics that occurred in a large category of
non-equilibrium and chaotic systems \cite{Wang3, Wang4}. Another
important feature of our FP equation is that the diffusion
coefficient is both $q$ and $t$ dependent.

This work was in part supported by the National Natural Science
Foundation of China (Grant Nos. 70401020, 70571027, 10647125, and
10635020) and the Ministry of Education of China (Grant No. 306022).


\begin{thebibliography}{s2}
\bibitem{Wang1} Wang Q A 2005 Chaos, Solitons \& Fractals 23 1253
\bibitem{Tsallis1} Tsallis C 1988 J. Stat. Phys. 52 479
\bibitem{Fokker} Fokker A D 1914 Ann. Physik 43 810
\bibitem{Planck} Planck M 1917 Sitzber. Preuss. Akad. Wiss. p324
\bibitem{Uhlenbeck} Uhlenbeck G E and Ornstein L S 1930 Phys. Rev. 36 832
\bibitem{Chandrasekhar} Chandrasekhar S 1943 Rev. Mod. Phys. 15 1
\bibitem{Wang&Uhlenbeck} Wang M C and Uhlenbeck G E 1945 Rev. Mod.
Phys. 17 323
\bibitem{Haken} Haken H 1975 Rev. Mod. Phys. 47 67
\bibitem{Schuss} Schuss Z 1980 Theory and Applications of Stochastic Differential
Equations (New York: Wiley)
\bibitem{Risken} Risken H 1984 The Fokker-Planck Equation (Berlin:
Springer-Verlag)
\bibitem{Shannon1} Shannon C E 1948 Bell System Technical Journal
27 379-423 \& 623
\bibitem{Tsallis2} Tsallis C 1994 New Trends in Magnetism, Magnetic
Materials and Their Applications (New York: Plenum) p451
\bibitem{Prato} Prato D 1995 Phys. Lett. A 203 165
\bibitem{Wang2} Wang Q A, Bangoupa S, Dzanguea F, Jeatsaa A, Tsobnanga F and Le
M\'ehaut\'e A 2008 Chaos, Solitons \& Fractals in press
\bibitem{Michael} Michael F and Johnson M D 2003 Physica A 324 359
\bibitem{Tsallis3} Tsallis C and Bukman D J 1996 Phys. Rev. E 54 2197
\bibitem{Wang3} Wang Q A 2004 Chaos, Solitons \& Fractals 19 639
\bibitem{Wang4} Wang Q A, Le M\'ehaut\'e A, Nivanen L and Pezeril M 2004 Physica A 340 117

\end{thebibliography}
\end{document}